\documentclass[10pt,journal,compsoc]{IEEEtran}
\usepackage{hyperref}
\usepackage{framed}
\usepackage{caption}
\usepackage{multicol}
\usepackage{subcaption}
\usepackage{balance}
\usepackage{mathtools}
\usepackage{seqsplit}
\usepackage{amsmath}
\usepackage{amssymb}
\usepackage{algorithmic}
\usepackage{array}
\usepackage{wrapfig}
\usepackage{graphicx}
\usepackage{times}
\usepackage{tikz}
\usepackage{graphicx}
\usepackage{amsfonts}
\usepackage{url}
\usepackage[T1]{fontenc}
\usepackage{times}
\usepackage{mathtools}
\usepackage{longtable}
\usepackage{xcolor}
\usepackage{setspace}
\usepackage{tabularx,tikz}
\usepackage{tikz-cd}
\usepackage{graphicx}
\usepackage{enumitem}
\usepackage[boxed,ruled,vlined,linesnumbered]{algorithm2e}
\usepackage[none]{hyphenat}
\usepackage{theorem}
\usepackage{csquotes}
\usepackage{tabularx,tikz}
\newcommand{\F}{\vspace*{\smallskipamount}}

\newcommand{\BBB}{\vspace*{-\bigskipamount}}
\usepackage{tikz-cd}
\usetikzlibrary{matrix,shapes,arrows,positioning,chains, calc}

\usetikzlibrary{shapes.arrows,chains}
\usetikzlibrary{backgrounds}
\tikzstyle{mybox} = [draw=white,   rectangle]
\tikzstyle{fancytitle} =[fill=red, text=white]
\newcommand\newcaptionstyle[2]{%
  \expandafter\ifx\csname caption@@#1\endcsname\relax
    \defcaptionstyle{#1}{#2}%
  \else
    \PackageError{caption}{Caption style `#1' already defined}{}%
  \fi}
\newcommand\defcaptionstyle[2]{%
  \@namedef{caption@@#1}{#2}}

\makeatletter
    \renewcommand*{\@fnsymbol}[1]{\ensuremath{\ifcase#1\or *\or \dagger\or \ddagger\or
       \mathsection\or \mathparagraph\or \|\or **\or \dagger\dagger
       \or \ddagger\ddagger \else\@ctrerr\fi}}
    \makeatother

\ifCLASSOPTIONcompsoc
  \usepackage[nocompress]{cite}
\else
  \usepackage{cite}
\fi
\ifCLASSINFOpdf
\else
\fi

\IEEEoverridecommandlockouts

\usepackage[mathlines,switch]{lineno}

\begin{document}
\title{Smart Home Survey on Security and Privacy\\
\author{\IEEEauthorblockN{Nisha Panwar\IEEEauthorrefmark{1},
		Shantanu Sharma\IEEEauthorrefmark{1},
		Sharad Mehrotra\IEEEauthorrefmark{1},
		\L ukasz Krzywiecki\IEEEauthorrefmark{2},
		Nalini Venkatasubramanian\IEEEauthorrefmark{1}}
	\IEEEauthorblockA{\IEEEauthorrefmark{1} University of California, Irvine, USA.}
	\IEEEauthorblockA{\IEEEauthorrefmark{2} Wroc\l aw University of Technology, Poland.}}}

\IEEEtitleabstractindextext{
\begin{abstract}
Smart homes are a special use-case of the Internet-of-Things (IoT) paradigm. Security and privacy are two prime concern
in smart home networks. A threat-prone smart home can reveal lifestyle and behavior of the occupants, which may be a significant
concern. This article shows security requirements and threats to a smart home and focuses on a privacy-preserving security model.
We classify smart home services based on the spatial and temporal properties of the underlying device-to-device and owner-to-cloud
interaction. We present ways to adapt existing security solutions such as distance-bounding protocols, ISO-KE, SIGMA, TLS, Schnorr,
Okamoto Identification Scheme (IS), Pedersen commitment scheme for achieving security and privacy in a cloud-assisted home area network.	
\end{abstract}


}
\maketitle

\IEEEdisplaynontitleabstractindextext
\IEEEpeerreviewmaketitle
\ifCLASSOPTIONcompsoc
\IEEEraisesectionheading{\section{Introduction}\label{section:introduction}}
\else

\fi

The rapid growth of IoT is assisting humans in many applications such as healthcare, transportation, entertainment, industrial appliances, sport, building management, and homes. Such IoT environments, while they provide unprecedented opportunities, they raise significant security and privacy concerns. This article focuses on smart homes and buildings in which a variety of devices interact over the local network.
Table~\ref{tab:ly} enlists the typical devices in a home area network (HAN) along with the underlying communication protocols. The device connectivity in the HAN can be visualized as a star topology, where devices connect to a central device controller. The device controller is directly connected to the home router that provides connectivity with the external traffic. The design of a HAN brings in many challenges, as follows:

\smallskip
\noindent\textbf{Device heterogeneity} encompasses different embedded hardware, operating systems, and user interfaces; e.g., HAN devices might be installed with different operating systems such as Tiny OS (open source), Contiki (open source), and RIOT (micro-kernel based).

\smallskip
\noindent\textbf{Communication heterogeneity} encompasses different transmission medium protocols; e.g., different IoT specific communication protocol adaptations such as Message Queuing Telemetry Transport (MQTT), Constrained Application Protocol (CoAP), User Datagram Protocol (UDP), and Transport Control Protocol (TCP).

\smallskip
\noindent\textbf{Technical expertise.} Often, smart homeowners are not engineers or security experts. Thus, most of the smart homeowners manually actuate device interaction or to own a fully autonomous home.

\smallskip
\noindent\textbf{Resource constraints and data collection.} Most of the HAN devices are lack of computational and storage power. In contrast, HAN devices are a continuous source of data streaming. Hence, one needs to store the data of all devices at one place (possibly the cloud), which imposes secure and privacy-preserving data collection and processing tasks.

\begin{figure}[h]
	\begin{center}
		\includegraphics[scale=0.7]{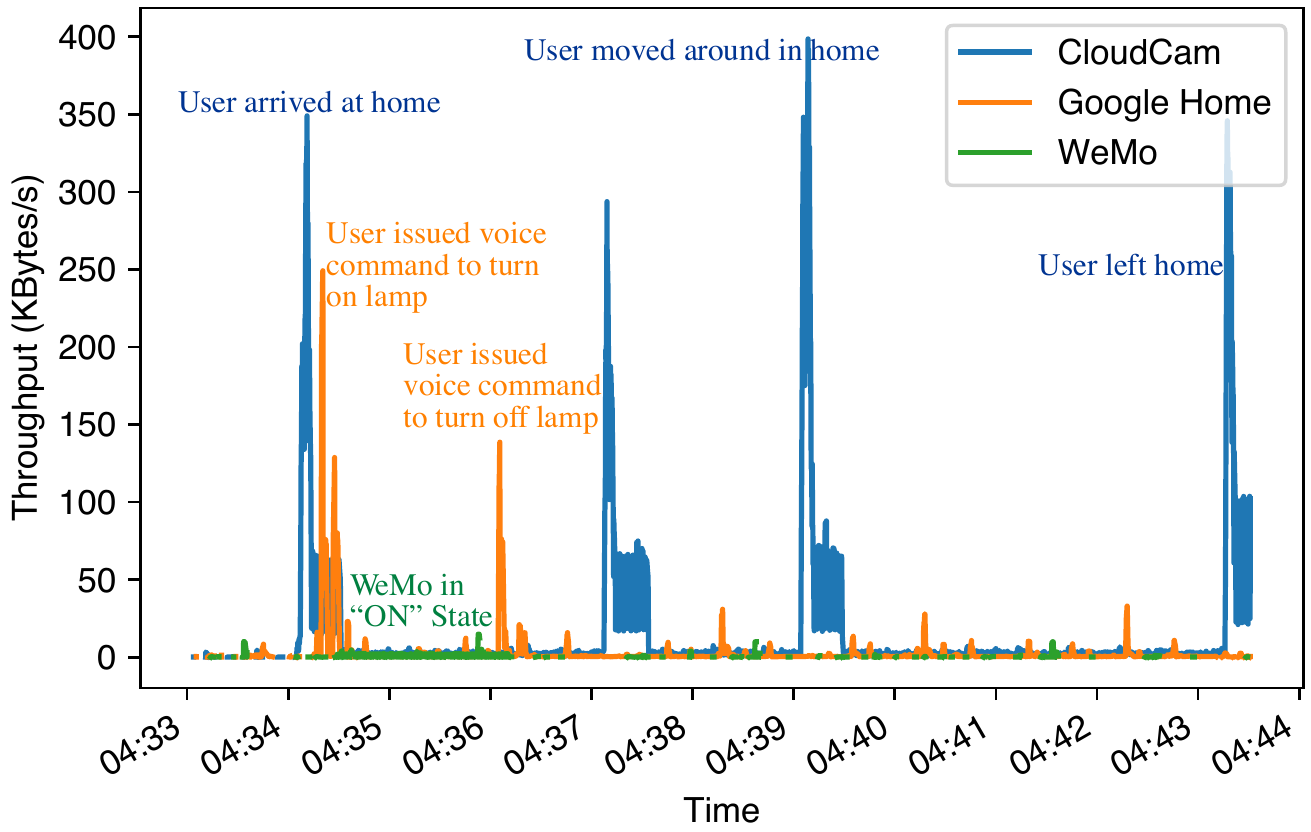}
	\end{center}
	\caption{Traffic activity in correlation with user activity.}
	\label{fig:peak}
\end{figure}

\smallskip
\noindent\textbf{Security and privacy.} The HAN is susceptible to passive monitoring over wireless channels, which may reveal potential information regarding user interaction, behavior, lifestyle or physical activity. We consider a privacy threat due to inference attacks such that a homeowner might loose the control over meta-information leakage about the {\em user activity} via {\em channel activity} or {\em device activity}. The figure~\ref{fig:peak} shows the channel activity for three home devices: CloudCam, Google Home, and Belkin WeMo. The CloudCam shows a peak in the channel activity (up to 400 KB/s traffic rate) as the user enters the home, moves inside the home or exits the home. The Google Home shows a peak in the channel activity (up to 250 KB/s traffic rate) whenever a user initiated a voice command for the light bulbs to turn on/off. Similarly, a bi-state WeMo switch peaks during the on state and creates a channel activity lesser than 20 KB/s.

Undoubtedly, the autonomous layout among the home devices would bring-in convenience. However, if the device interaction is not secure, it would give away the home control to anyone closely listening to the communication. At worse, even an adversary located at a remote location can hijack the device session by tampering with the external traffic. For example, a smart door lock receives encrypted commands through mobile applications over the wireless channel. If an adversary replays those commands used in a session between the door lock and the mobile application, then the door lock can be compromised through an identity mis-binding attack. In this article, we restrict ourselves to a secure HAN model for proximity-based communication. Below we discuss two use-cases to show the need of security and privacy in HANs beyond the naive solutions mentioned above.

\begin{table*}[t]
\begin{center}
    \begin{tabular}{|p{2cm}|p{7.5cm}|p{2.5cm}|p{4.5cm}|}
    \hline
\textbf{Usage} & \textbf{Example devices} &\textbf{Communication protocols} &\textbf{Scenarios} \\\hline

Voice-based smart assistants & Alexa, Amazon Echo, Lenovo Smart Assistant, Eufy Genie (far-field Alexa control), iHome iAVS 16, HTC U11,  & BLE, WiFi, ZigBee, Z-Wave & D2DWL (Alexa: what is the morning traffic status?) \\\hline

Healthcare & Zio XT Patch, HealthPatch MD, Sleep monitors & BLE, WiFi & D2DWL (Alexa: How much did I walk or sleep today?) \\\hline

Lightening & Lutron-Sivoia QS Shading System/Caseta Pro Dimmer, Philips-Hue Bloom/Lightstrip Plus, & BLE, WiFi, ZigBee, Z-Wave & D2DW, D2DWL (Alexa: Put the shades on enough intensity for reading/watching TV/eating.) \\\hline

Door locks & Schlage Connect Touchscreen Deadbolt, Ring Doorbell Pro & Cellular, WiFi, ZigBee, Z-Wave & D2DWL (Alexa: Who is at the door?) \\\hline

HVAC systems & ecobee4 Thermostat & BLE, WiFi, ZigBee, Z-Wave & D2DWL (Alexa: Is it more humid today?) \\\hline

Controllers & Amazon Echo, Amazon Echo Dot, ARRIS SURFboard router (1750Mbps wireless speed, 1.4Gb download speed, gigabit ethernet), ASUS Blue Cave (based on Intel), Belkin WeMo Mini,
Logitech Harmony Remote (use with Alexa, upto 15 devices) & WiFi, ZigBee, Z-Wave & D2DW, D2DWL (Alexa: Put lights on daylight bright, music on quite, microwave on grill mode, washing machine on dry mode.) \\\hline

Hygiene & SteriGrip self-cleaning door handles, Unico smartbrush   & WiFi, Cellular & D2DWL (Alexa: Restore my Unico timer and brush rotation speed.)  \\\hline

Remote surveillance & Logitech Circle 2, Amazon Echo Show & WiFi, Cellular & D2DW, D2DWL (Alexa: Show me who came at the door between 9am to 5pm?) \\\hline

Smart meters &  Itron, Elster, Landis+Gyr/Toshiba, Aclara, ABB and Sensus Metering
Systems& WiFi, ZigBee, Z-Wave & D2DW, D2DWL (Alexa: Tell me the peak power consumption on hours/days/weeks?) \\\hline

Smart dust & Vehicle keyfob, eco-cooking tools, smart gardening tools, and wild life saver & Bluetooth, WiFi & D2DWL (Alexa: Keep the smart dust in low-power mode when not in use.) \\\hline

Entertainment & Sonos Play: 1, Sonos Playbar, Amazon Fire TV, Sony XBR TV, DISH Hopper 3, & WiFi, ZigBee, Z-Wave & D2DW, D2DWL (Alexa: Play my morning mantra playlist.) \\\hline

\multicolumn{4}{|c|}{Notations: BLE: Bluetooth Low Energy, D2DWL: Device-to-Device Wireless, D2DW: Device-to-Device Wired.} \\\hline
\end{tabular}
\end{center}
\caption{Smart home devices and underlying communication protocols.}
\label{tab:ly}
\end{table*}


\smallskip
\noindent\textbf{Use-Case 1: Privacy.} HAN devices are dependent and connected such that controlling few devices would jeopardize the entire HAN. Furthermore, device traffic sniffing can lead to user activity inferences. The naive solution is to isolate the HAN from all types of external traffic entering the home gateway and internal traffic leaving the home gateway, though impractical. Another solution is to use {\em benign duping} such that the group of devices inside a home would simulate the owner activity, e.g., the owner is not in the home, yet the lights would turn on/off periodically as if the owner is at home. We provide privacy-preserving owner-to-cloud interaction such that the cloud can verify the owners' identity but cannot infer the original identity.



\smallskip
\noindent\textbf{Use-Case 2: Vandalism.} The door locks, camera surveillance, and motion detectors make a secure physical periphery of the home. However, if these devices communicate through a public channel, then an intruder might be able to first bypass the cyber-security of these devices, and then the physical periphery of the home. Hence, it is required to secure the HAN device connectivity for the overall secure automation.

This article studies security protocols for device-to-device (D2D) and owner-to-cloud (O2C) interactions in HANs. The device interaction is classified as per the proximity communication model that utilizes spatial and temporal attributes to perceive the security requirements. We provide a brief glance through existing security protocols for each of these proximity communication scenarios.


\smallskip\noindent\textbf{Related work.} Smart homes are a combination of different standards: \textit{IEEE P2413 Internet of Things} defines the Internet-enabled device communication with the assistance of  infrastructure. \textit{IEEE 1888.4 Green Smart Homes} defines the smart home paradigm from energy resource management perspective. \textit{IEEE 1547 Smart Grid Integration} defines how to connect distributed renewable energy resources with the smart grid. \textit{IEEE 2030} defines how to interoperate various smart grid technologies. \textit{IEEE 1901} defines an infrastructure using broadband over power line. \textit{IEEE 2302 Intercloud Interoperability} defines new means of intercloud connectivity. However, there is no standard security specification for smart home networks. 

Currently, the smart home communication consists of a set of autonomous devices capable of sensing and conducting actions. There exists a study about the social barrier into adoption of home networks~\cite{BaltaOzkan2013363}. User~\cite{goto} can also send commands or trigger a self-executing order~\cite{douzaR17} for these devices from a remote location. Overall, the smart home communication so far is visualized as a network component that executes commands based on contextual factors as well as overriding the contextual commands with the priority commands from the smart home owner. To the best of our knowledge, none of the work~\cite{firs,thir} highlight the significance of secure ordering inside smart homes much similar to non-digital homes. We highlight the presence of a pattern among smart home devices such that a partial ordering on device activity is observed on daily basis. Usually, devices perform actions during similar hours and along with similar logically neighbouring devices over the digital timeline in IoT homes. The authors in~\cite{thir} have presented a privacy-preserving traffic shaping scheme to mask the channel activity and thereby the device or user activity at the ISP level. According to the scheme, if the shaped traffic rate is lower than the device traffic then the packets are queued, and if the shaped traffic rate is higher than the device traffic then dummy packets are added to cover the original traffic rate variations. However, these techniques do not avoid the inferences on device activity pattern due to the straightforward binding between the channel activity and device activity. Our scheme decouples the channel activity from the device activity such that a communication activity over the channel at any given time cannot be coupled with a specific device activity or the user activity. 

There exists a number of IoT frameworks~\cite{doan,fram} based on a general model that includes IoT devices, backend cloud and a proxy gateway. These frameworks support a variety of IoT devices from the vendors such as Amazon, Samsung, Google, Philips Hue, Nest, Belkin and others, therefore, the connectivity requirements aka communication protocols vary in each of these framework scenario. Here, we provide a brief overview of these IoT frameworks and the security layer within:

\noindent\textit{Apple HomeKit:} This IoT framework is dedicated to smart home device connectivity. It leverages the connectivity for IoT home appliances and accessories through smartphone iOS apps. The iOS app {\em Home} allows the devices to join/leave home network, customize, and control the home environment. It must be noted that an owner can choose actions for IoT devices through Siri service in Homekit, however, it is still not possible to schedule the device actions in advance as we propose in this paper.  In the HomeKit architecture IoT devices connect to the platform either directly or through proxy gateways that supports ZigBee and Z-Wave communication protocols. However, the IoT devices that directly connects through {\em HomeKit} accessory protocol can communicate through LAN, WiFi or BLE instead of ZigBee and Z-wave transport protocols. For example, tvOS 10 supports the {\em Homekit} framework and acts like a hub for IoT home devices and a home owner can access these device from a remote location through iOS app. The security layer in {\em Homekit} includes {\em Perfect Forward Secrecy} (PFS) and secure communication over Transport Layer Security (TLS) or Datagram TLS with AES128-GCM, AES256 and SHA256. The PFS ensures that any future communication is secure and the leakage of long-term keys in future cannot reveal the sessions from past. In addition, the applications' access to home data is based on permission-model and the iOS system data is secure against memory-based attacks through {\em Address Space Layout Randomization} (ASLR) technique.

\noindent\textit{Amazon Web Service (AWS):} This IoT platform provides a ubiqutous connectivity between the IoT devices and the AWS cloud. The AWS architecture includes: (\textit{a}) {\em device gateway} providing connectivity among IoT devices and the cloud services through MQTT (Message Queue Telemetry Transport), SSL (Secure Socket Layer), TLS, Websockets and HTTP (Hyper Text Transfer Protocol) 1.1; (\textit{b}) {\em device shadows} maintaining a virtual replica of the original device also keep synchronizing the device state. In case a device is offline the device shadow retains the last visible state of the device and all pending upgrades can be restored once the device is online; (\textit{c}) {\em rule engine} providing a policy execution on the published data and transforming it into subscriber-appropriate format; (\textit{d}) {\em registry} maintaining the meta-level information (e.g., device name, identity, vendor, other attributes, etc) about connected devices. The security architecture of AWS includes authentication based on $X.509$ certificates, confidentiality through SSL/TLS based secure key exchange, access control through policy specification, forward secrecy through TLS cipher suites such as AES128-GCM-SHA256, ECDHE-ECDSA-AES128-GCM-SHA256, and, AES256-GCM-SHA384.

\noindent\textit{Samsung SmartThings:} This IoT platform is dedicated to smart home environments and appliance connectivity through mobile phone apps. The {\em SmartThing} framework is composed of a cloud backend, controller hub, mobile client app, and the IoT devices. The controller hub interacts with the home devices and the cloud services. The hub provides connectivity through several communication protocols such as ZigBee, Z-Wave, WiFi, and BLE. In addition, the cloud-connected devices utilize {\em OAuth/OAuth2} protocol for authentication and SSL/TLS for message transmission. In addition, the hub supports AES-128 bit encryption for all communication with ZigBee and Z-Wave enabled products.

\noindent\textit{Azure IoT Suite:} This IoT platform includes IoT devices, cloud services and hub to provide secure connectivity. The cloud is entitled to send commands and notifications for the IoT devices through the hub. The hub supports MQTT and HTTP protocols to enable this bi-directional connectivity. The security layer provides device authentication, access control and communication security. The device authentication is based on HMAC-SHA256 signed token along with the unique device identity. The access control and authorization is based on permission policies defined in the {\em Azure Active Directory}. The SSL/TLS protocol is used for secure handshake, mutual authentication, and session secrecy.

There exists other frameworks such as {\em IBM Watson} IoT platform, {\em Brillo/Weave} platform by Google, {\em Calvin} IoT platform by Ericsson, {\em ARM mbed} IoT platform, {\em Kura} IoT project by Eclipse, interested readers may refer~\cite{fram} for more details.

\begin{table*}[!h]
	\scriptsize
	\begin{center}
		\begin{tabular}{|p{1.5cm}|p{5cm}|p{5.9cm}|p{0.8cm}|p{1.6cm}|p{1.8cm}|}
			\hline
			\textbf{Attacks} & \textbf{Purpose} &\textbf{Scenario} &\textbf{Channel} & \textbf{Security compromise}  & \textbf{Techniques} \\\hline
			
			\multicolumn{6}{c}{\textbf{Existing passive attacks}} \\\hline
			
			Eavesdropping/ Side-channel & Scanning all the communication channels to know information of participants and their behavior. & D2DWL (e.g., sleep monitor activity reveals whether the user is awake or not?) & WiFi, ZigBee, Z-Wave,  & Availability, confidentiality & Traffic shaping, padding \\\hline
			
			Repeater-in-the -Middle, Replay & Amplification or retransmission of wireless signals from past & D2DW, D2DWL (e.g., vehicle-to-keyfob, where keyfob thinks it connects to a parked vehicle while a repeater in the middle apmlify the signal way too early and gets access to the vehicle ) & WiFi, ZigBee, Z-Wave & Authentication, integrity  & Out-of-band or multi-factor authentication \\\hline
			
			\multicolumn{6}{c}{\textbf{Existing active attacks}} \\\hline
			
			Jamming, evasion, spoofing & Overwhelms the network resources and makes them inaccessible to legit users & D2DWL (e.g., in the presence of traffic congestion at same frequency and shared bandwidth channel such that devices are unable to connect with the smart meter) & WiFi, ZigBee, Z-Wave & Availability & Topology and neighbor discovery \\\hline
			
			Man-in-the -Middle  & Actively engaging with two devices such that at least one of them cannot see the adversary in middle & D2DW, D2DWL (e.g., a mobile application connects with the paired door lock while actually connecting with a standby middle adversary and vice-versa) & LAN, WiFi, ZigBee, Z-Wave  & Authentication &  Identification \\\hline
			
			Identity misbinding & Active attack on two devices in interaction such that one of these authentic device establish key with the active adversary & D2DW, D2DWL (e.g., closing the door lock with a mobile application where mobile app connects with the door lock but the door lock connects with an active adversary and thereby follow the fabricated commands) & LAN, WiFi & Authentication & Identification \\\hline
			
			Repudiation & Ability to disown the communication transcripts or the identities used for communication & D2DW, D2DWL (e.g., Alice leaving no traces of her original identity while she opened the door lock with Bob's password) & LAN, WiFi, ZigBee, Z-Wave & Non-repudiation, identification & Signature or certificate-based identification  \\\hline
			
			Tampering & Ability to reveal or modify the secret messages & D2DW, D2DWL (e.g., Alice requesting smart meter to send power usage while she receives power usage of previous months added with the current usage) & LAN, WiFi, ZigBee, Z-Wave & Confidentiality, integrity & Escrowing, auditable proofs  \\\hline
			
			Denial of services (DoS) & Malicious users overwhelm the network resources and makes them inaccessible to other legitimate users & D2DWL (e.g., Alexa reacting first to the user who is closer and therefore, the user who is a little farther would not be able to reach Alexa and receive the response) & WiFi, ZigBee, Z-Wave & Availability &  Neighbor discovery \\\hline
			
			Reflection attacks & Initiator device ends up establishing a secure communication with itself & D2DW, D2DWL (e.g., keyfob seems synchronize and ping car window but car window does not seem to follow the command and open up) & LAN, WiFi, ZigBee, Z-Wave & Authentication, forward secrecy &  Multi-factor authentication \\\hline
			
			Routing table attacks & Aimlessly roaming messages in the network without a path validation mechanism & D2DW, D2DWL (e.g., home controller cannot find a few indoor devices on the network topology and therefore, cannot connect with them even if located very closely) &  LAN, WiFi, ZigBee, Z-Wave & Confidentiality, integrity & On-the-fly path validation  \\\hline
			
			Firewall piercing & Bypassing the security mechanism and the ability to establish a covert channel & D2DW, D2DWL (e.g., secret identification of the sender device based on a clock-skewness pattern in the outgoing TCP/ICMP messages) &  LAN, WiFi, ZigBee, Z-Wave & Authentication, integrity  & Digital watermarking, fingerprinting, intrusion detection \\\hline
			
			Destructive attacks & Randomly switching on/off the devices, disrupting indoor temperature, sound settings, breaking water outlets & D2DWL (e.g., ability to access the devices for maintenance and to disrupt the settings knowingly) & LAN, WiFi, ZigBee, Z-Wave & Authorization   & Privacy- preserving communication \\\hline
			
			Key compromise & Revealing the static/ephemeral key to impersonate user identity (e.g., Alice compromise the master key of digital-home crypto-key store and change all other keys) & D2DW, D2DWL (e.g., social engineering attacks to deceive the user into revealing the keys) & LAN, WiFi, ZigBee, Z-Wave &   Forward secrecy, backward secrecy, adaptive secrecy & Key escrow, secret key shares, ephemeral keys  \\\hline
			
			
			\multicolumn{6}{c}{\textbf{New attacks in smart environment}} \\\hline
			
			
			Application compromise & Installing a malware for man-in-middle, id-based or combination of attacks, compromising other connected devices and sensitive data, transferring fake data & D2DWL (e.g., Alice login to a bank website and during logout an active adversary blocks the finish message and proper session termination)  & All external traffic & Authentication, confidentiality, integrity & Memory partitioning, middleware approach \\\hline
			
			OS compromise & Exploit available memory/data of devices, unauthorized access, double-pricing  & O2C (e.g., Meltdown, Heartbleed and other information leakage attacks during process execution) & Trojans, malware & Authentication, confidentiality & Secure co-processors, Intel SGX \\\hline
			
			Cloud compromise & Attacks on clouds, virtual machines, communications between the HAN gateway and the cloud; revealing data/computations at cloud & O2C (e.g., Data access pattern attacks) & LAN, WiFi, ZigBee, Z-Wave & Authentication, confidentiality, integrity, non-repudiation, availability & Proxy- reencryption, homomorphic encryption \\\hline
			
			Location compromise & GPS spoofing, distance-attacks & D2DWL, O2C (e.g., using fabricated pseudorandom noise codes) & BLE, WiFi, Cellular& Pre-shared keys &  Anonymity, unlinkability, unobservability \\\hline
			
			
			Social networking & These attacks focus on breaking at least one device containing user sensitive data by fake apps, plug-ins, offers, click hijacking, botnets, and impersonation and then gradually increasing the overall influence, e.g., sybil attacks & D2DW, D2DWL (e.g., by stealing devices and pin codes) & LAN, WiFi, ZigBee, Z-Wave & Confidentiality & Plausible deniability, anonymity \\\hline
			
			Evil-twin & A fake access point mimicking an authentic Service Set Identifier (SSID) & D2DWL (e.g., a standalone access point deceiving users to connect through it and then launching other attacks such as packet sniffing) & WiFi & Authentication, confidentiality, integrity, availability & Wireless/device fingerprinting  \\\hline
			
			
		\end{tabular}
	\end{center}
	\caption{Existing and new security attacks in the HAN.}
	\label{tab:security attacks}
\end{table*}

\section{Security Threats and Goals}
\label{subsec:Attacks and Security Solutions}
Existing HANs are vulnerable to a variety of active and passive attacks.  Table~\ref{tab:security attacks} presents a taxonomy on these threats. Below we provide an overview of the basic security requirements necessary for the proximity-based communication scenarios in the HAN.

\begin{itemize}[noitemsep,nolistsep,leftmargin=0.1in]
  \item \textit{Authentication:} confirms identities of participants and provides an evidence to each participant as to whom they are communicating with. Digital signatures, secure identity transmission, physically-unclonable functions (PUF), and secure key exchanges are viable solutions for authentication.

  \item \textit{Authorization:} grants privileges to an authenticated user. Fine-grained policy management and different access-control methods are used to provide access-rights.

  \item \textit{Confidentiality:} hides the messages from an adversary and reveal to an authorized user only. Encryption, secret-sharing, traffic padding, zero-knowledge proofs, proof-of-knowledge, group signatures, pseudonym systems provide confidentiality.

  \item \textit{Integrity:} assures the correctness and consistency of the identity and messages. Cryptographic hashing, watermarking, holographic proofs, multi-party computations, timestamping, nonce, and sequence numbers are used to maintain integrity.

  \item \textit{Availability:} guarantees fair operations, resources, and services to an authorized user. Anomaly detection, firewalls, and special communication hardware preventing external malicious traffic to reach the network are techniques for achieving fairness.

  \item \textit{Non-repudiation:} prevents either a sender or a receiver from denying a transmitted message or a protocol execution. Digital signatures are commonly used to provide non-repudiation.

  \item \textit{Accountability or auditing:} is a process of bookkeeping each step at the sender, the receiver, or the network, so that a \emph{judge} or a participant can verify the transactions in future.

  \item \textit{Deniability:} The deniable communication empowers a prover to plausibly deny that a protocol instance was ever executed, for which the same prover was an active participant. Therefore, even if all protocol transcripts are stored for later analysis, it does not suffice as an identification-proof for a specific prover that participated during the protocol.



  \item \textit{Unlinkability:} A user might utilize multiple pseudonyms for authentication regarding different services, however, this must not yield a linking between any two pseudonyms used by the same homeowner for two different services.


  \item \textit{Non-transferability:} A homeowner has unique privileges over the HAN. For example, only one unique administrator account is the highest privileged on a standalone computer. Thus, the non-transferability over these privileges is crucial to retain the ownership. This property avoids \textit{credential forgery} and the user \textit{identity attacks}.

\item \textit{Forward secrecy:} In order to enable temporal security for cryptographic credentials perfect forward secrecy~\cite{karejo} is crucial. Forward secrecy guarantees that a session key derived from a long-term public-private key pair is secure even if one of the (long-term) private keys are compromised in the future. 

\item \textit{Privacy:} The privacy is inherent in establishing and maintaining the HAN because the most economic attack is to do passive learning and gain side-channel information without even breaking the lengthy keys and complex hash codes. Essentially, D2D communication should be able to protect the privacy through hiding: user activity, behavior pattern, device activity patterns. 
\end{itemize}




\section{Lightweight Cryptographic Solutions}
\label{sect:solutions}
This section presents ways to incorporate existing lightweight cryptographic solutions for HANs. The HAN consists of computational and storage inefficient devices (e.g., door locks, coffee machines, thermostats) that receive on/off commands \textit{vs} resource-abundant complex devices (e.g., smart meters, voice-based assistants) that additionally collect data, trigger alarms, inform the owner about device state or any resource shortage. These devices interact with each other using different communication protocols, as explained in Table~\ref{tab:ly}. Thus, the HAN would require a combination of security protocols based on symmetric keys, asymmetric keys, Authenticated Key Exchange (AKE), reactive authentication, out-of-band authentication, or privacy aware identification. Our selection of these adapted protocols guarantees security properties such as secrecy, integrity, confidentiality, authentication, authorization, accountability, forward secrecy, non-transferability, unlinkability, and plausible deniability.


\noindent\textit{Lightweight:} Our discussion below is limited to lightweight cryptographic protocols~\cite{somithree} to support secure D2D interactions in HAN wherein devices are connected to and controlled  by a central controller over a WiFi network. The protocols execute over cheap commodity hardware with limited processing power, memory, and communication bandwidth as is the case with current IoT devices. We note that the future may support a more decentralized device architecture wherein home devices may execute more autonomously (e.g., a washing machine (in order to reduce cost) may exploit the dynamic pricing model of resources such as water and electricity) and use data payment protocols using  cryptocurrency that require proof-of-mining and consume larger processor, memory-usage, and communication.

\subsection{Proximity Model}
The proximity communication model in a HAN can be envisioned on two dimensions: space and time, as follows:

\smallskip
\noindent\textbf{Timeline.} The notion of the timeline is to incorporate communication activity of every pair/group of devices under the same roof. In addition, each device shares its local timeline with a peer device. The time representation for device activation requires a dependency relation among those devices. Moreover, it would be easier to represent how much \textit{time} and \textit{energy resources} a device spends in communication with other devices. Consequently, these devices and groups can be merged into a single \textit{virtual node} as they stand adjacent on the timeline.


\smallskip
\noindent\textbf{Space.} The notion of space considers spatial aspects of digital devices in the HAN. The spatial arrangement of smart devices is another important dimension to create a \textit{virtual node}. In addition to the devices that spend most of the time in communication with a specific device, there exist devices that interact less frequently. However, these devices share the spatial locality.

The device interaction can be categorized based on \textit{space} and \textit{time}, as follows:

\begin{itemize}[noitemsep,nolistsep,leftmargin=0.1in]
	\item \textbf{Same Time and Same Space:} These are an indoor group of devices that share an identical timeline and the location. These devices would require wireless communication such as Bluetooth, infrared, wireless LAN, ZigBee, and Z-Wave. This scenario can be adapted with the existing out-of-band verification methods to guarantee multi-factor authentication. We briefly discuss the out-of-band verification protocols for these devices, e.g., distance-bounding.

	\item \textbf{Same Time and Different Space:} These are an outdoor group of devices that share the common timeline but different geographical locations. These devices would require wired or cellular (e.g., LTE) based connectivity. This scenario can be adapted with the existing secure key exchange protocols to guarantee per-session security. We briefly discuss the security protocols ISO-KE, SIGMA, and TLS for these devices.
	
	\item \textbf{Different Time and Different Space:} This use-case requires the homeowner interaction with a third-party service provider located at a distance. These interactions include service registration, activation, computation, and termination. This scenario can be adapted with privacy-preserving identification and access delegation protocols that assure security features (e.g., non-transferability, unlinkability, and deniability). These solutions provide control over the identification such that committed values can be revealed to a verifier in future and the right to reveal that secret information is uniquely held by the owner only. We briefly discuss Schnorr~\cite{beryyinluv} and Okamoto IS, Pedersen bit-commitment protocol, proxy re-encryption, and homomorphic encryption approaches for these scenarios.
	
	\item \textbf{Different Time and Same Space:} This use-case requires scheduling protocols~\cite{douzaR17} such that devices receive an actuation token, perform computation, produce results, and pass the token to next device in the queue. In~\cite{cody} we present a new verifiable delay scheme that allows an unpredictable amount of artificial delay before the command actuation at any home device. In particular, our solution provides improved privacy guarantees over naive solutions that require command pre-scheduling at these IoT devices.
\end{itemize}

\subsection{Key Generation}

The general methods for secure key management are key pre-sharing, key evolution, and PUF-enabled key-store. In key pre-sharing techniques, a set of potential keys are pre-shared with  devices. Each device shares a subset of pre-shared keys and the common intersection of those subsets is used to generate the future session keys.

The key evolution methods generate a symmetric key or a pair of asymmetric keys for each device. The symmetric key solutions require two devices to share a bi-directional static secret key that encrypts/decrypts messages at both devices. As an advantage, the symmetric key solutions require the minimum number of exponentiation per device than asymmetric key solutions. However, this computation efficiency comes with a vulnerability to key compromise attacks. In contrast, the asymmetric-key-based protocols provide additional security features such as non-repudiation, fewer number of keys to be stored, and the secure storage on insecure media. In addition, key evolution provides methods to frequently change the secret keys while the corresponding public key remains unchanged.

In PUF secured key-store, all secret keys and pre-shared keys are stored in a single key-storage, however, this key-storage is accessible only through a master secret key. Therefore, PUF can be used to secure these secret master keys without explicitly storing the master key on a storage media. Therefore, it avoids the \textit{key extraction attacks} through physically access to the device storage.

We assume a smart home scenario consists of $n$ home devices ($D_1,D_2, \ldots,D_n$) and an owner $O$. D2D communication (wired or wireless communication channel) must be augmented with secure sessions. A secure session key agreement requires that communicating parties share a long-term static secret and use a new ephemeral secret for each session. Also, a secure session must begin with the validation of static credentials, i.e, who has generated, distributed and revoked the authentic credentials. A trusted-third-party, called Public Key Infrastructure (PKI) or Key Generation Center (KGC), e.g., Kerberos system, is used to manage these master keys.

We classify device communication into three categories: device-to-device wireless (D2DWL) connectivity, device-to-device wired (D2DW) connectivity, and owner-to-cloud (O2C) connectivity. The classification of these use-cases is derived from time-space analogy defined earlier. 

%
\subsection{Device-to-Device Wireless (D2DWL) Connectivity}
The \textit{same time and same space} based proximity model is applicable in D2DWL, because the device interaction through a wireless channel in a close proximity is less vulnerable to security breaches than a device interaction at a remote location. Home to parking (vehicle/keyfob), home to irrigation controllers, home to garbage bins, and home to surveillance unit are examples of D2DWL. For securing D2DWL, a reactive authentication or out-of-band verification methods can be used. The reactive authentication method combines (\textit{i}) \emph{static attributes} (e.g., a secret key, hardware-based challenge-response verification using PUF, wireless fingerprinting, time-space localization --- triangulation, trilateration, multilateration, and distance-bounding~\cite{dacha}); and (\textit{ii}) \emph{dynamic attributes} (e.g., the user behavior or the anthropomorphic features including biometrics, gait, voice, and typing patterns) --- for generating a random challenge for which only an authentic entity can generate the paired response. We briefly explain the security adaptations based on the distance-bounding protocols.

\medskip
\noindent\textbf{Distance-bounding:} Location is a unique attribute for verifiable service access such as providing building access only when the user possesses right magnetic token and the user is close enough to the door. We chose distance-bounding protocol that verifies authentic credentials along with the proximity estimation over wireless channels. The distance-bounding protocols infer an upper bound on the distance between a sender and a receiver by measuring the round-trip time (RTT) of signals. In general, these protocols use $n$ rounds for accurate distance estimation by exchanging unique challenge-response pairs between a sender and a receiver. A sender forwards a unique challenge to an intended recipient within the proximity. Then, the receiver receives the challenge and computes the paired response using a trusted hardware or a shared hash function. The sender will receive this response and check the validity of response such that no older response has been replayed. Now, the sender can estimate the distance of the party sending the correct response by knowing (\textit{i}) RTT: the interval when the challenge was sent to recipient and the corresponding response was received, (\textit{ii}) measuring the time while radio signal traversing at the speed of light. Consequently, an estimate distance is computed such that no sender farther than the estimated distance could have transmitted the signals without incurring an additional delay over the current estimation.

\subsection{Device-to-Device Wired (D2DW) Connectivity}
The \textit{same time and different space} based proximity model is applicable in D2DW, because the wired communication is secure for long distance interactions in real-time. For example, device connectivity with the smart meters, smart grid, broadband over power line (\textit{IEEE 1901}) etc. In D2DW, the security can be achieved using either a symmetric key (e.g., Needham Schroeder symmetric key protocol) or asymmetric keys (e.g., AKE, non-interactive Diffie-Hellman, PKI based ISO-KE, SIGMA, TLS/SSL, NAXOS, HMQV, CMQV protocol). We present the asymmetric key based solutions, namely ISO-KE, SIGMA, and TLS protocols using Diffie-Hellman (DH) key exchange.

In Diffie-Hellman algorithm, all computations are done within a cyclic group $G=\langle g \rangle$ of a prime order $q$, where Computational Diffie-Hellman (CDH) assumption holds. The CDH assumption satisfies that the computation of a Discrete Logarithm (DL) function on public values $(g,g^a,g^b)$ is hard~\cite{deborah} within group $G$.

\medskip	
\noindent\textbf{Computational Diffie-Hellman.}	Let $\langle g \rangle$ be a cyclic group $G$ generated by an element $g$ of order $q$. There is no efficient probabilistic algorithm $\mathcal{A}_{\mathit{CDH}}$ that given $(g,g^a,g^b)$ produces $g^{\mathit{ab}}$, where $a$, $b$ are chosen as random group elements.


\begin{figure*}[!t]
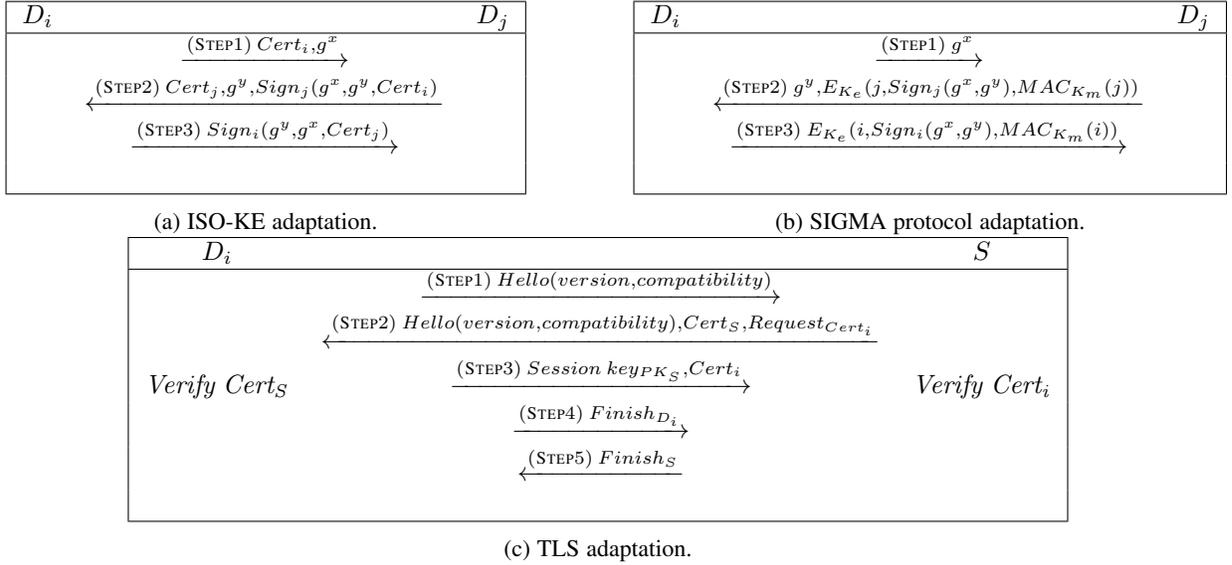

	\begin{center}
		\begin{minipage}{.48\linewidth}
			\centering
			\begin{tabular}{|c c c|}
				\hline
				$D_i$ &  & $D_j$ \\
				\hline
				& $\xrightarrow{(\textsc{Step} 1) \: Cert_i, g^x}$ & \\
				& $\xleftarrow{(\textsc{Step} 2) \: Cert_j, g^y, Sign_j(g^x,g^y, Cert_i)}$ & \\
				& $\xrightarrow{(\textsc{Step} 3) \: Sign_i(g^y,g^x, Cert_j)}$ & \\
				& & \\
				\hline
			\end{tabular}
			\subcaption{ISO-KE adaptation.}
			\label{fig:ISO-KE}
		\end{minipage}
		\begin{minipage}{.48\linewidth}
			\centering
			\begin{tabular}{|c c c|}
				\hline
				$D_i$ &  & $D_j$\\
				\hline
				& $\xrightarrow{(\textsc{Step} 1) \: g^x}$ & \\
				& $\xleftarrow{(\textsc{Step} 2) \: g^y, E_{K_e}(j, Sign_j(g^x,g^y), MAC_{K_m}(j))}$ & \\
				& $\xrightarrow{(\textsc{Step} 3) \: E_{K_e}(i, Sign_i(g^x,g^y), MAC_{K_m}(i))}$ & \\
				&  & \\
				\hline
			\end{tabular}
			\subcaption{SIGMA protocol adaptation.}
			\label{fig:SIGMA}
		\end{minipage}		
		
		\begin{minipage}{.98\linewidth}
			\centering
			\begin{tabular}{|c c c|}
				\hline
				$D_i$ &  & $S$\\
				\hline
				& $\xrightarrow{(\textsc{Step} 1) \: Hello (version, compatibility)}$ & \\
				& $\xleftarrow{(\textsc{Step} 2) \: Hello (version, compatibility), Cert_S, Request_{Cert_i}}$ & \\
				$\mathit{Verify \: Cert_S}$ & $\xrightarrow{(\textsc{Step} 3) \: Session \: key_{PK_S}, Cert_i}$ & $\mathit{Verify \: Cert_i}$  \\
				& $\xrightarrow{(\textsc{Step} 4) \: Finish_{D_i}}$ & \\
				& $\xleftarrow{(\textsc{Step} 5) \: Finish_S}$ & \\
				& & \\
				\hline
			\end{tabular}
			\subcaption{TLS adaptation.}
			\label{fig:TLS}
		\end{minipage}
	\end{center}
	\BBB
	\caption{ISO-KE, SIGMA, TLS protocols adaptation in D2DW scenario of a HAN.}
	\label{fig:ISO-KE, SIGMA, TLS protocols}
	\BBB
\end{figure*}

\medskip\noindent\textbf{ISO-KE protocol adaptation:} According to ISO-KE (Figure~\ref{fig:ISO-KE}), the sender device $D_i$ initializes a secure key exchange by sending its certificate, $\mathit{Cert}_i$, and a DH exponent, $g^x$, to a receiver, $D_j$ (\textsc{Step} 1). After the certificate identity verification, $D_j$ responds to $D_i$ with its own certificate, $\mathit{Cert}_j$, and its DH exponent, $g^y$. However, in order to avoid any misbinding/reflection attacks, $D_j$ concatenates the credentials of the intended recipient $D_i$, i.e., $\mathit{Sign}_j(g^x,g^y,\mathit{Cert}_i)$ (\textsc{Step} 2). $D_i$ responds with the signature on mutually agreed values, i.e., $\mathit{Sign}_i(g^x,g^y,Cert_j)$ (\textsc{Step} 3). The session key is derived from $g^{\mathit{xy}}$. However, note that the ISO-KE protocol does not support \textit{identity hiding}, since certificates are transferred in plaintexts.

\medskip\noindent\textbf{SIGMA protocol adaptation:} SIGMA supports sender and recipient  identity hiding feature unless the identity of the opposite device is successfully verified. In this protocol, the DH key exchange is authenticated through digital signature, and the device identity is encrypted using a freshly driven key ($K_e$) to protect the identity against eavesdropping.

When adapting SIGMA protocol in a HAN (Figure~\ref{fig:SIGMA}), a device $D_i$ sends DH exponent $g^x$ to an intended recipient $D_j$ (\textsc{Step} 1). On receiving $g^x$, $D_j$ computes its exponent $g^y$ and the corresponding session key $g^{\mathit{xy}}$. Then, $g^{\mathit{xy}}$ is used to generate three different and computationally independent keys, i.e., a session key ($K_s$), an encryption key ($K_e$), and a message authentication key ($K_m$). $D_j$ forward an encrypted message with the exponent $g^y$, signed exponents ($g^x,g^y$), and the Message Authentication Code (MAC) over identity $j$ (\textsc{Step} 2). The protocol terminates after $D_i$ verifies the signed exponents and the identity from $D_j$ and computes the corresponding session key $g^{\mathit{xy}}$. In addition, $D_i$ forwards an encrypted message with the signed exponents ($g^x,g^y$) and the MAC over identity $i$ (\textsc{Step} 3).


As an advantage, this protocol has the minimal number of message exchange than any key exchange protocol does to prevent replay attacks. Figure~\ref{fig:SIGMA} shows a 3-round version of SIGMA protocol with the prover identity protection as a required feature. Similarly, there also exists a 4-round version of SIGMA that provides verifiers' identity protection. Whenever the indoor home network devices do not want to reveal the identity before verifying the identity of a peer device/service in long-distance communication, then the SIGMA protocol would be preferred over ISO-KE protocol.

\medskip\noindent\textbf{Transport Layer Security (TLS) protocol adaptation:}
TLS is a widely accepted standard for Internet security (Figure~\ref{fig:TLS}). TLS handshakes are based on a pre-defined sequence of phases: mutual authentication, random secret exchange, session key establishment, and finish. The handshake between a device $D_i$ and a server $S$ starts by sending $\mathit{Hello}$ messages that include supported range of cryptographic standards called cipher-suite (\textsc{Step} 1 and \textsc{Step} 2). Also, the mutual authentication is accomplished through authority signed certificates in a $\mathit{Certificate \: Exchange}$ message (\textsc{Step} 2 and \textsc{Step} 3). $D_i$ sends session key to server $S$ along with its certificate in (\textsc{Step} 3). Also, $D_i$ sends a finish message to acknowledge the beginning of ciphered communication (\textsc{Step} 4). Similarly, $S$ sends a finish message to acknowledge the beginning of ciphered communication (\textsc{Step} 5). Afterward, $D_i$ and $S$ switch to the symmetric key encryption-based communication using the recently established session key to encrypt and decrypt the message exchange.

\smallskip\noindent\textit{Comparisons.} SIGMA protocol is preferred for the applications where privacy is crucial along with an end-to-end session security. SIGMA protocol provides identity-hiding features which are neither part of ISO-KE nor TLS. In addition, SIGMA protocols can be chosen either for sender identity-protection (3-rounds) or for receiver identity-protection (4-rounds).

\subsection{Owner-to-Cloud (O2C) Connectivity}
\label{subsec:oc}
The \textit{different time and different space} based proximity model is considered in O2C, because the data produced by devices in a home is often stored at a different space, particularly, the public cloud at a different time, due to some processing at the home and the network transmission delay.

Some home devices send data to cloud, and this data needs to be accessed in a privacy-preserving manner. Though encrypting the data at the cloud,  accessing encrypted data at the cloud can reveal users' privacy of the user. For example, if the user data is accessed by a cancer hospital, then by access-patterns and background knowledge, an adversary can reveal medical conditions of the user without knowing the encrypted data and users' identity. Hence, the ease of using cloud services comes with threats to the data and user privacy, and it brings two main challenges:



\begin{enumerate}
  \item \textit{Secure storage and query processing} to prevent the cloud to learn user's activities, through usage-patterns or query processing on the data. Hence, data storage and query processing at the cloud must be cryptographically secure using techniques, such as encryption or secret-sharing.

  \item \textit{Secure authentication} provides a way to securely authenticate the homeowner at the cloud. For this purpose, we illustrate Schnorr and Okamoto IS executing between the homeowner and the cloud. Despite having a secure connection between the cloud and the homeowner, the cloud can still reveal the proof of communication that can be avoided using secure commitment protocols. Bit-commitment protocols are based on \textit{commitment-before-knowledge} paradigm and can be used for a private proof of identity possession in HAN. A bit-commitment protocol incorporates a \textit{commit} phase and a \textit{reveal} phase such that committed value is revealed only if the DL condition satisfies. We illustrate an adaptation of Pedersen commitment protocol in HAN.
\end{enumerate}

\medskip\noindent\textbf{Schnorrs' Identification Scheme:} The scheme is based on the intractability of DL problem. According to the protocol, an owner $O$ as identity prover selects a secret DH exponent $x$ and releases a corresponding public value $X$ for the verifier cloud $C$. Consequently, the verifier cloud returns a challenge $\mu$ for the identity proving owner. Now the prover generates a combined response $\rho$ such that it is computationally hard to compute $\rho$ without possessing the knowledge of $x$. Furthermore, there exist multiple variations to regular Schnorr like IS. For example, one possible way~\cite{keser} is to replace $\rho$ with $\hat{\rho}$, i.e., $\hat{g}^{x+a*{\mu}}$ where $\hat{g}={\mathcal{H}(X|\mu)}$ is the new generator such that the IS remains robust against the \textit{ephemeral key leakage}.

\noindent\textit{Common inputs} are ($p,q,g,G$) where $p,q$ are large prime numbers, $g$ is the generator of order $q$ in group $G$. \\
\noindent\textit{Keygen(pk,sk)}: $sk=a \: {\in} \: \mathbb{Z}_{q}$ and $pk=g^a=A$. The protocol $\Gamma(O,C)$ steps for the prover $O(a,A)$ and the verifier $C(A)$ are given below:

\begin{itemize}[noitemsep,nolistsep,leftmargin=0.1in]
	\item Prover selects $x \in \mathbb{Z}_{q}$  at random in Schnorr's group and computes $X=g^x \in G$.
	\item Prover sends $X$ to the verifier.
	\item Verifier computes random $\mu \: {\in} \: \mathbb{Z}_{q}$ and sends $\mu$ to prover.
	\item Prover computes response $\rho=x+a*{\mu}$ and sends $\rho$ to verifier.
	\item Verifier accepts the proof of commitment if $g^{\rho}=g^{x+a*{\mu}}={(g^x)}*{{(g^a)}^{\mu}}=\mathit{XA^{\mu}}$.
\end{itemize}

\medskip\noindent\textbf{Okamoto Identification Scheme:} Okamoto IS~\cite{satos} is based on DL problem and provides a proof of long-term secret key possession at the prover. The owner $O$ possess secret keys ($a_1,a_2$) and prove it to the verifier cloud using an ephemeral secret ($x_1,x_2$). Consequently, the verifier cloud returns a challenge $\mu$ for the identity proving owner. Now the prover generates a combined response ($\rho_1,\rho_2$) such that it is computationally hard to compute ($\rho_1,\rho_2$) without possessing the knowledge of ($x_1,x_2$).

\noindent\textit{Common inputs} are ($p,q,g,G$) where $p,q$ are large prime numbers, $g$ is the generator of order $q$ in group $G$. Also, $g_1,g_2 \in G$ where ${log_{g_1}}{g_2}$ is unknown.

\noindent\textit{Keygen(pk,sk)}: The prover $O$ knows the secret key $sk_O=(a_1,a_2)$ with the public key $pk_O$ as $A=({g_1}^{a_1}*{g_2}^{a_2})$. The protocol $\Gamma(O,C)$ steps for prover $O(sk_O,pk_O)$ and verifier $C(pk_O)$ are given below:

\begin{itemize}[noitemsep,nolistsep,leftmargin=0.1in]
	\item Prover choose $(x_1,x_2) \in \mathbb{Z}_{q}$ at random in Schnorr's group and then compute $X=g^{x_1}*g^{x_2}$. 
	\item Prover sends $X$ to verifier. 
	\item Verifier computes a random challenge $\mu \: {\in} \: \mathbb{Z}_{q}$ and send to prover. 
	\item Prover sends $\rho_1=x_1+{\mu}*a_1$ and $\rho_2=x_2+{\mu}*a_2$ to verifier. 
	\item Verifier completes the successful identification ${g_1}^{\rho_1}*{g_2}^{\rho_2}={X}*A^{\mu}$.
\end{itemize}

\medskip\noindent\textbf{Pedersens' Commitment Scheme:} The Pedersen scheme~\cite{chour} is an unconditionally hiding scheme such that it is secure against an unbounded adversary. We further illustrate an adaptation of Pedersens' bit-commitment protocol in the scope of home networks where an owner communicates to a third party.

\noindent\textit{Common inputs} are ($p,q,g,G$) where $p,q$ are large prime numbers, $g$ is the generator of order $q$ in group $G$. Also, $g_1,g_2 \in G$ where ${log_{g_1}}{g_2}$ is unknown to the prover. The protocol $\Gamma(O,C)$ steps are below. The prover $O(m)$ commits a message $m$ and verifier $C(c)$ verifies the commitment $c$.

\begin{itemize}[noitemsep,nolistsep,leftmargin=0.1in]
	\item \textit{Commit:} Prover randomly selects $r \in \mathbb{Z}_{q}$ and send $c={g_1}^r*{g_2}^m$ to the verifier.
	\item \textit{Reveal:} Verifier receives the ($m,r$) and output $m$ if ${g_1}^r*{g_2}^m=c$.
\end{itemize}


\smallskip\noindent\textit{Comparisons.} Schnorr and Okamoto schemes are used for proving the identity, i.e., the knowledge of the secret key/keys, but Okamoto scheme is secure in stronger model (active adversary) at the cost of higher computational complexity (more exponentiations). Schnorr scheme is secure with passive adversary observing the protocol messages, i.e., the honest-verifier model. Both schemes can also be adapted to provide provably secure digital signatures through Fiat-Shamir transformation~\cite{shafi}. Unfortunately, both Schnorr and Okamoto ISs do not withstand the \textit{ephemeral key leakage attacks} such that security of long-term keys depends on the security of ephemeral keys. For the immunity against such attacks there exists modified Schnorr and modified Okamoto that require additional computational complexity. Interestingly, Pedersen's commitment scheme is computationally binding and unconditionally hiding. Therefore, Pedersen's scheme is useful as a sub-procedure for the applications that require security against an unbounded adversary.

\begin{table*}[t]
	\centering
	\begin{tabular}{ | p{2.4cm} | p{2.2cm} | p{2cm} | p{2.4cm} | p{1.7cm} | p{4.5cm} | }
		\hline
		\textbf{Protocols} & \textbf{Direct iteration cost} & \textbf{Certificate preprocessing} & \textbf{Exponentiation cost (per party)} & \textbf{Out-of-band verification} & \textbf{Purpose} \\ \hline
		ISO-KE~\cite{jade} & 3 rounds & Yes & 1 & No & Secure key exchange (D2DW) \\ \hline
		SIGMA~\cite{rypto} & 3 rounds & No & 1 & No & Secure key exchange (D2DW)  \\ \hline
		TLS~\cite{kellog} & $>$3 rounds & Yes & 1 & No & Secure key exchange (D2DW, D2DWL) \\ \hline
		Schnorr IS~\cite{beryyinluv} & 4 rounds & No & 1 & No & Identification (O2C) \\ \hline
		Okamoto IS~\cite{chour} & 3 rounds & No & 2 & No & Identification (O2C) \\ \hline
		Pedersen commitment~\cite{chour} & 3 rounds & No & 2 & No & Zero-knowledge proof of knowledge (O2C) \\ \hline
		Distance- bounding~\cite{dacha} & $>$4 rounds & No & 1 & Yes & Authentication (D2DWL) \\ \hline
	\end{tabular}
	\F
	\caption{Protocol comparison.}
	\label{tab:analyze}
\end{table*}

\medskip\noindent\textbf{Computations at the cloud:} For sharing the data with a third user, proxy re-encryption-based access delegation is applicable such that a proxy key is generated by the homeowner and proxy storage node would re-encrypt the data with this key that could only be decrypted using the secret key of the third user. The re-encryption key ${\mathit{rk}_{(a,b)}}$ is generated using a method $\mathit{Keygen}(\mathit{sk}_a,\mathit{pk}_a,\mathit{sk}_b,\mathit{pk}_b)$, where $a$ is the delegator (e.g. homeowner) and $b$ is the delegatee (e.g., cloud, law agency or auditor) $\mathit{sk}$ is the secret key, and $\mathit{pk}$ is the public key. This access delegation could be either unidirectional or bidirectional meaning that the data access rights can either be for a specific delegatee or mutually shared between both the delegator and the delegatee.

Proxy re-encryption converts the ciphered data into non-transferable ciphertext such that a colluding proxy and the delegatee cannot transfer access rights to a third party. Therefore, either the delegator/homeowner can access the data or the delegatee/cloud can access the data, even if the proxy colludes with any third party. For example, a proxy re-encryption scheme based on Elgamal assumes a group $G=\langle g \rangle$ of a prime order $q$. Also, the long-term public key is computed as $g^x$ where $x \in \mathbb{Z}_{q}$ is the secret key and $g$ is the group generator. A message $m$ is encrypted as $c_a=((g^x)^r,m \: g^r)$ for a random $r \in \mathbb{Z}_{q}$ by access delegator $a$. Re-encryption keys are generated as ${rk_{(a,b)}}$ = ${y/ x}$ mod $q \in \mathbb{Z}_{q}$ where $x$ and $y$ are the long-term secret key of party $a$ and $b$. Now the re-encryption process at the intermediate proxy node would compute exponentiation by using re-encryption key ${y/ x}$ mod $q$ on the first part of ciphertext $((g^x)^r)$ as $((g^{xr})^{(y/ x)},m \: g^r)$=$(g^{yr},m \: g^r)$. The access delegatee $b$ then generate original message $m=c_{a,2}{c_{a,1}}^{-(1/ y)}$.

For computing on encrypted data, many techniques are proposed in the literate and can be used. For example, homomorphic encryption allows computations over encrypted data without decrypting the data. Homomorphic encryption provides data confidentiality during data aggregation process; however, it is significantly slow. Here, we briefly explain Pailler's homomorphic encryption scheme that provides additive property, such that product of two encrypted text would be same as addition of the same text in plaintext. According to Pailler's homomorphic encryption scheme: $$c=g^mr^n \: \textnormal{mod} \: n^2$$ $$m=(c^{\lambda} \: \textnormal{mod} \: n^2 \: - \: 1)/n \: {\mu} \: \textnormal{mod} \: n$$

\noindent where $c$ is ciphertext, $m$ is a message in plaintext, $r$ is a random number $r \in \mathbb{Z}_{q}$, ($n,g$) is a public key and ($\lambda,\mu$) is a private key. Overall, neither proxy re-encryption nor homomorphic encryption scheme requires the secret key for data computation or data access. Therefore, both of these approaches are useful in providing the secure computation on encrypted data.

\section{Performance}
\label{performance}

Table~\ref{tab:analyze} compares protocols on four different criteria.  The second column represents the direct iteration cost that determines the communication complexity. Furthermore, online interaction with the trusted authority for each interaction is time-consuming and vulnerable to congestion. Therefore, these adapted schemes omit the need of an online assistance from a trusted-third-party for every round of authentication. The third column shows whether the certificate pre-processing is required before the protocol execution or not? The fourth column shows the  computational complexity, i.e., the required number of exponentiations at each party during the protocol execution. The fifth column shows that only distance-bounding approaches require an out-of-band communication to cross-verify the identity over a wireless channel by measuring its tentative distance. Finally, the last column represents the interaction model (D2D or O2C) where these selected protocols are applicable.

Our implementations have been created using Python 3 with \textit{Charm Crypto} library \cite{kinyel}.
We tested several proof of concept implementations: (\textit{i}) on the NIST-approved elliptic curve \textit{prime192v1} \cite{dssnist};
(\textit{ii}) in  Schnorr groups, with 1024-bit safe primes (e.g., $p = 2q + 1$). Average \textit{computational time} for each protocol has been measured by running 1000  executions of each protocol. 
 The results are presented in Table~\ref{tab:perf}. The message exchange does not impose a latency overhead and the message latency is unrelated to the computations. Thus, each protocol implementation is created as a single program 
 and the computations of parties are interweaving.

\begin{table}[h]
	\centering
	\begin{tabular}{|l|r|r|}
		\hline
		\multicolumn{1}{|c|}{\textbf{Protocols}} & \multicolumn{1}{c|}{\textbf{EC prime192v1}} & \multicolumn{1}{c|}{\textbf{Integer Group}} \\ \hline
		ISO KE&                                  3.899 ms &                                   7.352 ms \\ \hline
		SIGMA&                                  5.134 ms &                                   10.799 ms \\ \hline
		Schnorr IS&                                   1.126 ms &                                    2.439 ms \\ \hline
		Okamoto IS&                                  1.969 ms &                                   4.154 ms \\ \hline
		Pedersen commitment&                                  1.898 ms &                                   2.552 ms \\ \hline
	\end{tabular}
	\F
	\caption{Execution times for different protocol versions.}\label{tab:perf}
\end{table}





\section{Conclusion}
The article studies security protocols for a proximity-based communication model in HANs. Broadly, the HAN model consists of device-to-device and owner-to-cloud interactions. The proximity-based communication model incorporates \textit{space} and \textit{time} as two dimensions. The suitability of these security protocols is guided by these two dimensions. We provide a comparison and evaluation for these security protocols. Subsequently, we find that the computation latency overhead for these protocols is comparable to wireless signal speed as a fraction of second, therefore, is readily applicable in HAN settings.



\bibliographystyle{ieeetr}
{\small \bibliography{RelatedWork-arxiv}}

\begin{thebibliography}{10}

\bibitem{BaltaOzkan2013363}
N.~Balta-Ozkan, R.~Davidson, M.~Bicket, and L.~Whitmarsh, ``Social barriers to
  the adoption of smart homes,'' {\em Energy Policy}, vol.~63, pp.~363 -- 374,
  2013.

\bibitem{goto}
D.~N. Kalofonos and S.~Shakhshir, ``Intuisec: A framework for intuitive user
  interaction with smart home security using mobile devices,'' in {\em 2007
  IEEE 18th International Symposium on Personal, Indoor and Mobile Radio
  Communications}, pp.~1--5, 2007.

\bibitem{douzaR17}
S.~M. D'Souza and R.~Rajkumar, ``Time-based coordination in geo-distributed
  cyber-physical systems,'' in {\em 9th {USENIX} Workshop on Hot Topics in
  Cloud Computing, HotCloud 2017}, 2017.

\bibitem{firs}
N.~Apthorpe, D.~Reisman, and N.~Feamster, ``Closing the blinds: Four strategies
  for protecting smart home privacy from network observers,'' {\em CoRR},
  vol.~abs/1705.06809, 2017.

\bibitem{thir}
N.~Apthorpe, D.~Reisman, S.~Sundaresan, A.~Narayanan, and N.~Feamster, ``Spying
  on the smart home: Privacy attacks and defenses on encrypted iot traffic,''
  {\em CoRR}, vol.~abs/1708.05044, 2017.

\bibitem{doan}
T.~T. Doan, R.~Safavi-Naini, S.~Li, S.~Avizheh, M.~V. K., and P.~W.~L. Fong,
  ``Towards a resilient smart home,'' in {\em Proceedings of the 2018 Workshop
  on IoT Security and Privacy}, pp.~15--21, 2018.

\bibitem{fram}
M.~Ammar, G.~Russello, and B.~Crispo, ``Internet of things: A survey on the
  security of {IoT} frameworks,'' {\em Journal of Information Security and
  Applications}, vol.~38, pp.~8 -- 27, 2018.

\bibitem{karejo}
H.~Krawczyk, ``Perfect forward secrecy,'' in {\em Encyclopedia of Cryptography
  and Security}, pp.~457--458, 2005.

\bibitem{somithree}
P.~Kumar, A.~Gurtov, J.~Iinatti, M.~Ylianttila, and M.~Sain, ``Lightweight and
  secure session-key establishment scheme in smart home environments,'' {\em
  IEEE Sensors Journal}, vol.~16, no.~1, pp.~254--264, 2016.

\bibitem{beryyinluv}
C.-P. Schnorr, ``Efficient identification and signatures for smart cards,'' in
  {\em Proceedings of the 9th Annual International Cryptology Conference on
  Advances in Cryptology}, pp.~239--252, 1990.

\bibitem{cody}
N.~Panwar, S.~Sharma, G.~Wang, S.~Mehrotra, and N.~Venkatasubramanian,
  ``Verifiable round-robin scheme for smart homes,'' in {\em Proceedings of the
  Ninth ACM Conference on Data and Application Security and Privacy}, CODASPY
  '19, pp.~49--60, 2019.

\bibitem{dacha}
S.~Brands and D.~Chaum, ``Distance-bounding protocols,'' in {\em Advances in
  Cryptology EUROCRYPT: Workshop on the Theory and Application of Cryptographic
  Techniques}, pp.~344--359, 1994.

\bibitem{deborah}
W.~Diffie and M.~Hellman, ``New directions in cryptography,'' vol.~22,
  pp.~644--654, 2006.

\bibitem{keser}
{\L}.~Krzywiecki, ``Schnorr-like identification scheme resistant to malicious
  subliminal setting of ephemeral secret,'' in {\em Innovative Security
  Solutions for Information Technology and Communications: 9th International
  Conference, SECITC}, pp.~137--148, 2016.

\bibitem{satos}
T.~Okamoto, ``Provably secure and practical identification schemes and
  corresponding signature schemes,'' in {\em Advances in Cryptology},
  pp.~31--53, 1993.

\bibitem{chour}
D.~Chaum and T.~P. Pedersen, ``Wallet databases with observers,'' in {\em
  Proceedings of the 12th Annual International Cryptology Conference on
  Advances in Cryptology}, pp.~89--105, 1993.

\bibitem{shafi}
M.~Abdalla, J.~H. An, M.~Bellare, and C.~Namprempre, ``From identification to
  signatures via the fiat-shamir transform: Minimizing assumptions for security
  and forward-security,'' in {\em Advances in Cryptology --- {EUROCRYPT}},
  pp.~418--433, 2002.

\bibitem{jade}
``Iso/iec is 9798-3, entity authentication mechanisms, part 3: Entity
  authentication using asymmetric techniques,'' 1993.

\bibitem{rypto}
H.~Krawczyk, ``{SIGMA:} the 'sign-and-mac' approach to authenticated
  diffie-hellman and its use in the ike-protocols,'' in {\em Advances in
  Cryptology, 23rd Annual International Cryptology Conference}, pp.~400--425,
  2003.

\bibitem{kellog}
T.~Dierks and E.~Rescorla in {\em The Transport Layer Security (TLS) Protocol,
  Version 1.2}, 2008.

\bibitem{kinyel}
J.~A. Akinyele, C.~Garman, I.~Miers, M.~W. Pagano, M.~Rushanan, M.~Green, and
  A.~D. Rubin, ``Charm: a framework for rapidly prototyping cryptosystems,''
  {\em J. Cryptographic Engineering}, vol.~3, no.~2, pp.~111--128, 2013.

\bibitem{dssnist}
``{NIST, C:} digital signature standard available at url:
  \url{http://nvlpubs.nist.gov/nistpubs/FIPS/NIST.FIPS.186-4.pdf},''

\end{thebibliography}

\ifCLASSOPTIONcaptionsoff
  \newpage
\fi

\end{document}